\documentclass[12pt,a4paper,twoside]{article}
\usepackage[T1]{fontenc}
\usepackage{textcomp}
\usepackage[english]{babel}

\usepackage{mathpazo}
\usepackage[scaled]{helvet}

\usepackage{amssymb}

\newtheorem{theorem}{Theorem}
\newtheorem{lemma}{Lemma}
\newtheorem{corollary}{Corollary}
\newtheorem{property}{Property}
\newtheorem{definition}{Definition}
\newenvironment{proof}{\par\noindent\upshape\textbf{Proof.}~}{\qed\medskip}

\newcommand\pref[1]{\mathbf{pref}(#1)}
\newcommand\qed{{\hspace*{\fill}\mbox{$\square$}\par}}

\title{A Polynomial-Time Algorithm for the Automatic
  {Baire} Property} 
\author{\itshape{}\bfseries{}Ludwig
  Staiger\thanks{email: \ttfamily{}\textbf{staiger@informatik.uni-halle.de}}\\
  Martin-Luther-Universit\"at\ Halle-Wittenberg\\Institut f\"ur Informatik
  \\von-Seckendorff-Platz 1, D--06099 Halle (Saale), Germany} 
\date{}

\begin{document}
\maketitle{}

\thispagestyle{empty}

\begin{abstract}
  A subset of a topological space possesses the Baire property if it can be
  covered by an open set up to a meagre set. For the Cantor space of infinite
  words Finkel introduced the automatic Baire category where both sets, the
  open and the meagre, can be chosen to be definable by finite automata. Here
  we show that, given a Muller automaton defining the original set, automata
  for the resulting open and meagre sets can be constructed in polynomial
  time. The construction is based on the strongly connected subsets and the
  condensation graph of the automaton graph.

  Since the constructed sets are of simple topological structure, it is
  possible to construct not only Muller automata defining them but also the
  simpler B\"uchi automata. To this end we give, for a restricted class of
  Muller automata, a conversion to equivalent B\"uchi automata with an at most
  quadratic increase of the size.
\end{abstract}


\newpage{}

{\tableofcontents}

\bigskip{}

\section{Introduction}
\label{sec:int}

A subset of the Cantor space possesses the Baire property if it differs from
an open set only by a small set, that is, by a meagre set.  In the papers
\cite{LATA20/finkel,ijfcs21/finkel} the automatic Baire property as a special
case of the ordinary Baire property was introduced, and it was shown that
every automaton definable language of infinite words possesses not only the
Baire property but also the automatic Baire property.  To this end Finkel
constructed from a given Muller automaton two Muller automata accepting the
involved open and meagre set, respectively. Due to the simpler topological
structure of the open and meagre sets in Finkel's definition, we can
characterise them also by B\"uchi automata.

The aim of this paper is to present a polynomial time algorithm for the
automatic Baire property. For any given Muller automaton $\mathcal{A}$
accepting a regular $\omega$-language we construct in polynomial time two
Muller and two B\"uchi automata for the required $\omega$-languages.

Although the conversion, if possible, from Muller to B\"uchi automata may
result in an exponential blow-up (cf. \cite{dlt19/boker}), we show that in our
case it can be performed in polynomial time. To this end we prove that the
resulting automata can be kept small, thus allowing for an application of
standard polynomial-time graph algorithms.

\subsection{Notation}
\label{sec.notat}
In this section we introduce the notation used throughout the paper. By
$\mathbb{N} = \{ 0,1,2,\ldots\}$ we denote the set of natural numbers. Its
elements will be usually denoted by letters $i,\dots,n$. Let $X$ be a finite
alphabet of cardinality $|X| = r\ge 2$. Then $X^*$ is the set of finite words
on $X$, including the \textit{empty word} $e$, and $X^\omega$ is the set of
infinite strings ($\omega$-words) over $X$.  Subsets of $X^*$ will be referred
to as \textit{languages} and subsets of $X^\omega$ as
\textit{$\omega$-languages}.

For $w\in X^*$ and $\eta\in X^*\cup X^\omega$ let $w \cdot{}\eta$ be their
\textit{concatenation}.  This concatenation product extends in an obvious way
to subsets $W \subseteq X^*$ and $B\subseteq X^*\cup X^\omega$. For a language
$W$ let $W^* := \bigcup_{i \in \mathbb{N}} W^i$, and
$W^\omega:=\{w_1\cdots w_i\cdots: w_i\in W\setminus \{e\}\}$ be the set of
infinite strings formed by concatenating non-empty words in $W$.  Furthermore,
$|w|$ is the \textit{length} of the word $w\in X^*$ and $\pref B$ is the set
of all finite prefixes of strings in $B\subseteq X^*\cup X^\omega$.  We shall
abbreviate $w\in \pref{\{\eta\}}\ (\eta\in X^*\cup X^\omega)$ by
$w\sqsubseteq \eta$.

We assume the reader to be familiar with the basic facts of the theory of
regular languages and finite automata. We postpone the definition of
regularity for $\omega$-languages to Section~\ref{sec.regular}. For more
details on $\omega$-languages and regular $\omega$-languages see the book
\cite{PP04} or the survey papers \cite{handbook/St97,Th90}.

As usual, for a set $M$, we denote by $2^{M}:= \{M': M'\subseteq M\}$ its
\emph{powerset}.

\subsection{The Cantor space}
\label{sec.cantor}

We consider $X^\omega$ as a topological space (Cantor space). The
\emph{closure} (smallest closed set containing $F$) of a subset
$F \subseteq X^\omega, \mathcal{C}(F)$, is described as
$\mathcal{C}(F) := \{\xi : \pref{\{\xi\}} \subseteq \pref F\}$. The \emph{open
  sets} in Cantor space are the $\omega$-languages of the form
$W\cdot X^\omega$. Countable unions of closed sets are referred to as
\emph{$\Sigma_{2}$-sets}, their complements as \emph{$\Pi_{2}$-sets}.

Next we recall some topological notions.  As usual, an $\omega$-language
$F\subseteq X^{\omega}$ is \emph{dense in $X^{\omega}$} if
$\mathcal{C}(F)=X^{\omega}$. This is equivalent to $\pref F=X^{*}$. An
$\omega$-language $F\subseteq X^{\omega}$ is \emph{nowhere dense in
  $X^{\omega}$} if its closure $\mathcal{C}(F)$ does not contain a non-empty
open subset. This property is equivalent to the fact that for all
$w\in \pref F$ there is a $v\in X^{*}$ such that $w\cdot v\notin \pref
F$. Moreover, a subset $F\subseteq X^{\omega}$ is \emph{meagre} or of
\emph{first Baire category} if it is a countable union of nowhere dense sets.

\bigskip

\section{Automata and Regularity}
\subsection{Regular $\omega$-languages and finite automata}
\label{sec.regular}

As usual we call a language $W\subseteq X^{*}$ \emph{regular} if there is a
finite (deterministic) automaton $A = (X; S; s_0; \delta)$, where $S$ is the
finite set of states, $s_{0}\in S$ is the initial state and
$\delta:S\times X\to S$ is the transition function\footnote{We use the same
  symbol $\delta$ to denote the usual extension of the function $\delta$ to
  $S\times X^{*}$.}, such that $W=\{w: \delta(s_{0};w)\in S'\}$ for some fixed
set $S'\subseteq S$.

An $\omega$-language $F \subseteq X^\omega$ is called \emph{regular} provided
there are a finite (deterministic) automaton $A = (X; S; s_0; \delta)$ and a
table $\mathcal{T}\subseteq \{Z: Z\subseteq S\}$ such that for
$\xi\in X^\omega$ it holds $\xi\in F$ if and only if
$\mathrm{Inf}(A;\xi)\in \mathcal{T}$ where
$\mathrm{Inf}(A;\xi):= \bigcap_{w\sqsubset\xi}\{\delta(s_{0},v): w\sqsubset
v\sqsubset\xi\}$ is the set of all states $s\in S$ through which the automaton
$A$ runs infinitely often when reading the input $\xi$. Observe that
$Z= \mathrm{Inf}(A;\xi)$ holds for a subset $Z\subseteq S$ and some
$\xi\in X^{\omega}$ if and only if
\begin{enumerate}
\item there is a word $u\in X^{*}$ such that $\delta(s_{0};u)\in Z$, and
\item for all $s,s'\in Z$ there are non-empty words $w,v\in X^{*}$ such that
  $\delta(s,w)=s'$ and $\delta(s',v)=s$.
\end{enumerate}
Such sets were referred to as \emph{essential sets} \cite{ic/Wag79} or
\emph{loops} \cite{SelWa08},\cite[Section~5.1]{handbook/St97}, and
$\mathrm{LOOP}_{\mathcal{A}}= \{\mathrm{Inf}(\mathcal{A};\xi):\xi\in
X^\omega\}$ is the set of all loops of an automaton $\mathcal{A}$.  The
$\omega$-language
$L(\mathcal{A},\mathcal{T}) = \{\xi : \mathrm{Inf}(\mathcal{A};\xi)\in
\mathcal{T}\}$ is the union of all sets $\{\xi : \mathrm{Inf}(A; \xi) = Z\}$
where $Z\in \mathcal{T}$. The pair $(\mathcal{A},\mathcal{T})$ is usually
called a Muller automaton.

In particular, $\{\xi : \mathrm{Inf}(A; \xi) = Z\}$ and
$\{\xi : \mathrm{Inf}(A; \xi) = Z'\}$ are disjoint for $Z\ne Z'$.  Thus the
following holds.
\begin{lemma}\label{l.boolop}
  Let $\mathcal{A} = (X; S; s_0; \delta)$ be a deterministic automaton and
  $\mathcal{T},\mathcal{T}'\subseteq 2^{S}$ be tables, and let $\mathbf{op}$
  be a Boolean set operation. Then
  $L(\mathcal{A},\mathcal{T})\ \mathbf{op}\ L(\mathcal{A},\mathcal{T}')=
  L(\mathcal{A},\mathcal{T}\,\mathbf{op}\,\mathcal{T}')$. Moreover, for
  $\mathcal{T},\mathcal{T}'\subseteq 2^{S}$ we have
  $L(\mathcal{A},\mathcal{T})\subseteq L(\mathcal{A},\mathcal{T}')$ if and
  only if
  $\mathcal{T}\cap \mathrm{LOOP}_{\mathcal{A}}\subseteq\mathcal{T}'\cap
  \mathrm{LOOP}_{\mathcal{A}}$.
\end{lemma}

We are going to represent $F=L(\mathcal{A},\mathcal{T})$ by languages derived
from the automaton $\mathcal{A}$.  {As in
  \cite{bulleatcs/St98,dmtcs/St15} refer to a word $v\in X^*, v\ne e,$ as
  \emph{$(s; Z)$-loop completing} if and only if
  \begin{enumerate}
  \item $\delta(s, v) = s$ and $\{\delta(s, v') : v'\sqsubseteq v \} = Z$, and
  \item $\{\delta(s, v') : v'\sqsubseteq v'' \} \ne Z$ for all proper prefixes
    $v''\sqsubset v$ with $\delta(s, v'') = s$.
  \end{enumerate}}

Denote by $V_{(s;Z)}$ the set of all $(s; Z)$-loop completing words, and by
$U_s:=\{w:\delta(s_{0},w)=s\}$ the set of all words leading from the initial
state $s_{0}$ to the state $s\in S$. Both languages are regular and
constructible from the finite automaton $\mathcal{A} = (X; S; s_0; \delta)$.
Moreover, $V_{(s;Z)}$ is prefix-free and
$\pref{V_{(s,Z)}^\omega}=\{w:\forall w'(w'\sqsubseteq w\to \delta(s,w')\in
Z)\}$.

We obtain the following (cf. with \cite[Lemma~3]{bulleatcs/St98}).
\begin{lemma}\label{l.decomp}
  Let $\mathcal{A} = (X; S; s_0; \delta)$ be a finite automaton, and let
  $\mathcal{T}$ be a table. Then
  \begin{equation}\label{eq.decomp}
    L(\mathcal{A},\mathcal{T})=
    \bigcup_{Z\in\mathcal{T}}\bigcup_{s\in Z} U_{s}\cdot
    V_{(s;Z)}^\omega.  
  \end{equation} 
\end{lemma}
Thus every regular $\omega$-language has the form
$\bigcup_{j=1}^{\ell}W_{j}\cdot V_{j}^\omega$ where $W_{j}, V_{j}$ are regular
languages (see \cite{Bu60,PP04,handbook/St97} or \cite{Th90}). The converse is
also true, that is, if $W\subseteq X^{*}$ and $F, E\subseteq X^\omega$ are
regular then also $W^\omega, W\cdot{}E$ and $E\cup F$ are regular
$\omega$-languages. Note, however, that the representation of
Eq.~(\ref{eq.decomp}) is finer, since it splits a regular $\omega$-language
$F=\bigcup_{j=1}^{\ell}W_{j}\cdot V_{j}^\omega$ into parts
$U_{s}\cdot V_{(s;Z)}^\omega, i\in\{1,\dots,n\},$ where, additionally, the
languages $V_{(s;Z)}$ are prefix-free.

\subsection{Loops, strongly connected components and density}
\label{sec.loops}

We consider the loop structure of an automaton.  For $Z_1, Z_2\subseteq S$ we
write \mbox{$Z_1\mapsto Z_2$} if $Z_{1}\ne Z_{2}$ and there exists an
$s\in Z_1$ and a $w\in X^*$ such that $\delta(s,w)\in Z_2$.

The maximal w.r.t. ``\ $\subseteq$\ '' loops are the \emph{strongly connected
  components}\\
$\mathrm{SCC}_{\mathcal{A}}:= \{Z: Z\in \mathrm{LOOP}_{\mathcal{A}}\wedge
\forall Z'(Z'\in \mathrm{LOOP}_{\mathcal{A}}\to Z'\subseteq Z\vee Z'\cap
Z=\emptyset)\}$
of the automaton (multi-)graph of $\mathcal{A}$. They are the
vertices of the condensation graph of $\mathcal{A}$. For vertices
$Z,Z'\in \mathrm{SCC}_{\mathcal{A}}$ the relation $\mapsto$ is anti-symmetric
and transitive, thus a partial order. Its maximal w.r.t. ``\,$\mapsto$\,''
elements are the \emph{terminal strongly connected components}
$\mathrm{SCC}_{\mathcal{A}}^{t} := \{Z: \forall Z'(Z'\in
\mathrm{SCC}_{\mathcal{A}}\wedge Z\ne Z'\to Z\not\mapsto Z')\}$.

The following properties are immediate.
\begin{property}\label{pro.scc}
  Let $\mathcal{A} = (X; S; s_0; \delta)$ be a finite automaton and
  $Z\in \mathrm{SCC}_{\mathcal{A}}$.
  \begin{enumerate}
  \item If $z,\delta(z,w)\in Z$ then $\delta(z,w')\in Z$ for all
    $w'\sqsubseteq w$.\label{pro.scc1}
  \item If $z\in Z$ and $Z\in \mathrm{SCC}_{\mathcal{A}}^{t}$ then
    $\delta(z,w)\in Z$ for all $w\in X^{*}$.\label{pro.scc2}
  \end{enumerate}
\end{property}
\pagebreak{}

We have the following relation to the density of the
$\omega$-languages $V_{s,Z}^{\omega}$.
\begin{theorem}\label{th.termloop}
  Let $\mathcal{A}=(X; S; s_0; \delta)$ be an automaton and
  $Z\in \mathrm{LOOP}_{\mathcal{A}}$.
  \begin{enumerate}
  \item If $Z\notin \mathrm{SCC}_{\mathcal{A}}^{t}$ then $V_{s,Z}^{\omega}$ is
    nowhere dense.\label{th.termloop2}
  \item If $Z\in \mathrm{SCC}_{\mathcal{A}}^{t}$ then $V_{s,Z}^{\omega}$ is
    dense in $X^{\omega}$, that is,
    $\mathcal{C}(V_{s,Z}^{\omega})=X^{\omega}$.\label{th.termloop1}
  \end{enumerate}
\end{theorem}
\begin{proof}First, observe that
  $\pref{V_{s,Z}^{\omega}}=\{w: \delta(s,w)\in Z\}$.
  
 \ref{th.termloop2}. Let $w\in \pref{V_{s,Z}^{\omega}}$ and let
  $\delta(s',u)\notin Z$ for some $s'\in Z$ and $u\in X^{*}$. Consider
  $v\in X^{*}$ such that $\delta(s,wv)=s'$. Since
  $\delta(s,wvu)\notin Z$ we have $wvu\notin\pref{V_{s,Z}^{\omega}}$.
  
  \ref{th.termloop1}. If $Z\in \mathrm{SCC}_{\mathcal{A}}^{t}$ then
  $\delta(s,w)\in Z$ for every $w\in X^{*}$. Thus
  $\pref{V_{s,Z}^{\omega}}=X^{*}$, that is,
  $\mathcal{C}(V_{s,Z}^{\omega})=X^{\omega}$.
\end{proof}
As a consequence of Theorem~\ref{th.termloop} we have the following two lemmata.
\begin{lemma}\label{l.meagre}
  If $\mathcal{T}\cap \mathrm{SCC}_{\mathcal{A}}^{t}=\emptyset$ then
  $L(\mathcal{A},\mathcal{T})$ is of first Baire category.
\end{lemma}
\begin{proof}
  If $F\subseteq X^{\omega}$ is nowhere dense then also $w\cdot F$ is nowhere
  dense. Now the assertion follows with Lemma~\ref{l.decomp} and
  Theorem~\ref{th.termloop}.\ref{th.termloop2}.
\end{proof}
\begin{lemma}\label{l.open}
  If $Z\in \mathrm{SCC}_{\mathcal{A}}^{t}$ then
  $L(\mathcal{A},2^{Z})=\{\xi: \mathrm{Inf}(\mathcal{A}; \xi) \cap Z\ne
  \emptyset\}= \{w: w\in X^{*}\wedge \delta(s_{0}, w)\in Z\}\cdot X^{\omega}$
  is open in $X^{\omega}$.
\end{lemma}
\begin{proof}In view of Property~\ref{pro.scc}.\ref{pro.scc2}, $z\in Z$ and
  $Z\in \mathrm{SCC}_{\mathcal{A}}^{t}$ imply $\delta(z,w)\in Z$. Thus
  $Z\in \mathrm{SCC}_{\mathcal{A}}^{t}$ and
  $ \mathrm{Inf}(\mathcal{A}; \xi) \cap Z\ne \emptyset$ yield
  $\mathrm{Inf}(\mathcal{A}; \xi)\subseteq Z$.  Consequently,
  $\mathrm{Inf}(\mathcal{A},\xi)\subseteq Z$ when
  $\{\delta(s_{0}, w):w\sqsubset\xi\}\cap Z\ne\emptyset$.
\end{proof}
In particular, for a subset
$\mathcal{U}\subseteq \mathrm{SCC}_{\mathcal{A}}^{t}$ and its downward closure
$\mathcal{T_U}:=\{Z': \exists Z(Z\in\mathcal{U}\wedge Z'\subseteq Z)\}$, we can simplify an
automaton $\mathcal{A}=(X,S,s_{0},\delta)$ by merging the elements of the
terminal strongly connected components $Z\in \mathrm{SCC}_{\mathcal{A}}^{t}$
into a single state.
\begin{corollary}\label{c.open}Let $\mathcal{A}:=(X,S,s_{0},\delta)$ and
  $\mathcal{U}\subseteq \mathrm{SCC}_{\mathcal{A}}^{t}$. Then
  $L(\mathcal{A},\mathcal{T_U})= L(\mathcal{A'},\mathcal{T'})$ where
  $\mathcal{A'}:=(X,S',s_{0}',\delta')$ with
  \begin{eqnarray*}
    S'&:=& \bigl(S\setminus \bigcup\,\{Z:Z\in
           \mathrm{SCC}_{\mathcal{A}}^{t}\}\bigr)\cup\{s_0\}\cup \{s_{Z}:Z\in\mathrm{SCC}_{\mathcal{A}}^{t}\}\,,\\ 
    s_{0}'&:=&s_{0},\\
    \delta'(s,x)&:=&\left\{
                     \begin{array}{l@{,}l}
                       \delta(s,x)&\mbox{ if }\delta(s,x)\in S\setminus \bigcup\,\{Z:Z\in
                                    \mathrm{SCC}_{\mathcal{A}}^{t}\}, and\\[5pt]
                       s_{Z}&\mbox{ if }Z\in \mathrm{SCC}_{\mathcal{A}}^{t}\mbox{ and }\delta(s,x)\in
                          Z\mbox{ or }s=s_{Z}.
                     \end{array}\right.\\[3pt]
  \mbox{and }\mathcal{T'}&:= &\bigl\{\{s_Z\}:
  Z\in\mathcal{T_{U}}\cap\mathrm{SCC}_{\mathcal{A}}^{t}\bigr\}.
  \end{eqnarray*}
\end{corollary}

\begin{proof}By the Lemmata~\ref{l.open} and \ref{l.boolop}
  $L(\mathcal{A},\mathcal{T_U})= \bigcup_{Z\in \mathcal{U}}\{w:
  \delta(s_{0},w)\in Z\}\cdot X^{\omega}$. The construction of $\mathcal{A'}$
  shows, that for $w\ne e$ we have $\delta(s_{0},w)\in Z$ if and only if
  $\delta'(s_{0},w)=s_Z$. Now apply Lemma~\ref{l.open} to $\mathcal{A'}$.
\end{proof}
\subsection{The automatic Baire property}
\label{sec.Baire}
In this section we are going to prove an automatic version of the result
stating that every Borel (and even every analytic) set has the Baire property.
\begin{definition}\label{def.Baire}
  A subset $F\subseteq X^{\omega}$ has the \emph{Baire property} if there is
  an open set $E\subseteq X^{\omega}$ such that their symmetric difference
  $F\ \Delta\ E$ is of first Baire category.
\end{definition}
An important result of descriptive set theory is the following result, see
\cite{book/kuratowski66,Oxtoby80}.
\begin{theorem}
  Every Borel set of the Cantor space has the Baire property.
\end{theorem}
In \cite{LATA20/finkel,ijfcs21/finkel} an automatic version of the above
theorem is proved. We ﬁrst give the deﬁnition.
\begin{definition}[Automatic Baire property]\label{def.ABaire}
  A subset $F\subseteq X^{\omega}$ fulfils the \emph{Automatic Baire property}
  if there are a regular $\omega$-languages $E, F'$ such that
  $F\ \Delta\ E\subseteq F'$, where $E$ is open and $F'$ is a $\Sigma_{2}$-set
  of first Baire category.
\end{definition}

Then the following holds.
\begin{theorem}[\cite{LATA20/finkel,ijfcs21/finkel}]\label{th.finkel}
  Every regular $\omega$-language fulfils the Automatic Baire property.
\end{theorem}

For the purposes of our paper we give a proof. In \cite[Corollary 6]{St26} it
is shown that the requirement on $F'$ to be a $\Sigma_{2}$-set in
Definition~\ref{def.Baire} can dropped.

\begin{proof}
  For an automaton $\mathcal{A}:=(X,S,s_{0},\delta)$ and a table
  $\mathcal{T}\subseteq 2^{S}$ we have
  \begin{equation}\label{eq.fin}
    \begin{array}{rcl} \mathcal{T}&=& (\mathcal{T}\cap \mathrm{SCC}_{\mathcal{A}}^{t})\cup (\mathcal{T}\setminus
                                      \mathrm{SCC}_{\mathcal{A}}^{t})\subseteq (\mathcal{T}\cap
                                      \mathrm{SCC}_{\mathcal{A}}^{t}) \cup (2^{S}\setminus
                                      \mathrm{SCC}_{\mathcal{A}}^{t})\\[5pt]
                                  &\subseteq& \bigl\{Z': \exists
                                              Z(Z\in \mathcal{T}\cap
                                              \mathrm{SCC}_{\mathcal{A}}^{t}\wedge
                                              Z'\subseteq Z) \bigr\}\cup
                                              (2^{S}\setminus 
                                              \mathrm{SCC}_{\mathcal{A}}^{t}).
    \end{array}
  \end{equation}
  Since
  $\mathcal{T}\cap \mathrm{SCC}_{\mathcal{A}}^{t}=\bigl\{Z': \exists Z(Z\in
  \mathcal{T} \cap\mathrm{SCC}_{\mathcal{A}}^{t}\wedge Z'\subseteq
  Z)\bigr\}\cap \mathrm{SCC}_{\mathcal{A}}^{t}$, we have
  $\mathcal{T}\ \Delta\ \bigl\{Z': \exists Z(Z\in \mathcal{T}\cap
  \mathrm{SCC}_{\mathcal{A}}^{t}\wedge Z'\subseteq Z)\bigr\}\subseteq
  2^{S}\setminus \mathrm{SCC}_{\mathcal{A}}^{t}$.
Then Lemma~\ref{l.boolop} yields
    $L\bigl(\mathcal{A},\mathcal{T}\bigr)\ \Delta\
    L\bigl(\mathcal{A}, \bigl\{Z': \exists Z(Z\in \mathcal{T}\cap
    \mathrm{SCC}_{\mathcal{A}}^{t}\wedge 
    Z'\subseteq Z)\bigr\}\bigr)\subseteq
    L\bigl(\mathcal{A},2^{S}\setminus
    \mathrm{SCC}_{\mathcal{A}}^{t}\bigr).$
  According to Lemmata~\ref{l.open} and \ref{l.meagre} the $\omega$-languages
  $L\bigl(\mathcal{A}, \bigl\{Z': \exists
  Z(Z\in\mathrm{SCC}_{\mathcal{A}}^{t}\wedge Z'\subseteq Z)\bigr\}\bigr)$ and
  $L\bigl(\mathcal{A},2^{S}\setminus
  \mathrm{SCC}_{\mathcal{A}}^{t}\bigr)=X^{\omega}\setminus
  L\bigl(\mathcal{A},\mathrm{SCC}_{\mathcal{A}}^{t}\bigr)$ are open and of
  first Baire category, respectively.
\end{proof}
\section{Transformation of Muller to B\"uchi Automata}
\label{sec.transf}

In this section we consider the \emph{B\"uchi acceptance} of
$\omega$-automata. An $\omega$-lan\-guage $F\subseteq X^{\omega}$ is B\"uchi
accepted by an automaton $\mathcal{A}=(X; S; s_0; \delta)$ and a set of states
$T\subseteq S$ if
$F=\{\xi:\mathrm{Inf}(\mathcal{A},\xi)\cap T\ne \emptyset\}$.

It is well known \cite[Theorem~4.2]{La69} (cf. also
\cite{PP04,handbook/St97,Th90} or \cite{ic/Wag79}) that an $\omega$-lan\-guage
$F\subseteq X^{\omega}$ is accepted by a finite deterministic B\"uchi
automaton if and only if for every deterministic finite automaton
$\mathcal{A}$ and table $\mathcal{T}$ such that $F=L(\mathcal{A},\mathcal{T})$
the table $\mathcal{T}$ is \emph{upwards closed} with respect to
$\mathcal{A}$, that is, for $Z\in \mathcal{T}\cap \mathrm{LOOP}_{\mathcal{A}}$
and $Z'\in \mathrm{LOOP}_{\mathcal{A}}$ the inclusion $Z\subseteq Z'$ implies
$Z'\in\mathcal{T}$. This is equivalent to the fact that the $\omega$-language
$F\subseteq X^{\omega}$ is the complement of a regular $\Sigma_{2}$-set. In
particular, the tables
$\bigl\{Z': \exists Z(Z\in\mathrm{SCC}_{\mathcal{A}}^{t}\wedge Z'\subseteq
Z)\bigr\}$ and $\mathrm{SCC}_{\mathcal{A}}^{t}$ in the proof of
Theorem~\ref{th.finkel} are upwards closed.

Whether a translation from a Muller automaton $(\mathcal{A},\mathcal{T})$ to a
B\"uchi automaton $(\mathcal{A}',T)$ is polynomial, is unknown
(cf. \cite{dlt19/boker}). In case when the table $\mathcal{T}$ satisfies a
certain condition we obtain a quadratic increase in the number of states. This
implies that the obtained B\"uchi automaton has only a polynomial increase in
size.

\begin{theorem}\label{th.maxloop}
  Let $\mathcal{A}=(X; S; s_0; \delta)$ be an automaton and
  $\mathcal{T}\subseteq 2^{S}$ be a table such that
  $\mathcal{T}\cap\mathrm{LOOP}_{\mathcal{A}}\subseteq
  \mathrm{SCC}_{\mathcal{A}}$. Then there is a deterministic B\"uchi automaton
  $(\mathcal{A}';T)$ with $O(|S|^{2})$ states such that
  $L(\mathcal{A},\mathcal{T})=L(\mathcal{A}';T)$.
\end{theorem}
\begin{proof}
  Since all loops in $\mathcal{T}$ are strongly connected components of the
  automaton graph, they are pairwise disjoint. Let
  $\mathcal{T}\cap \mathrm{LOOP}_{\mathcal{A}} = \{Z_{1},\dots,Z_{n}\}$. For
  the purposes of the proof we assume that
  $z_{1}^{(i)},z_{2}^{(i)},\dots,z_{\kappa_{i}}^{(i)}, \kappa_{i}=|Z_{i}|,$ be
  a fixed ordering of $Z_{i}$.

  Define $\hat\mathcal{A}=(X; \hat S; \hat s_0; \hat\delta)$ as follows.
  \begin{eqnarray}
    \hat S&:=&\textstyle (S\times\{0\})\cup  \bigcup\nolimits_{i=1}^{n}
               (Z_{i}\times \{1,\dots, \kappa_{i}\}),\\
    \hat s_{0}&:=& (s_{0},0),\\
    T&=& \bigl\{(z_{\kappa_{i}}^{(i)},\kappa_{i}):i\in\{1,\dots,n\}\wedge z\in Z_{i}\bigr\},
  \end{eqnarray}
  for $s\in S$, and $x\in X$
  \begin{equation}\label{eq.A1}
    \hat\delta((s,0),x):= \left\{
      \begin{array}{l@{,}lr}
        (\delta(s,x),0)&\mbox{ if }\delta(s,x)\notin \bigl\{z_{1}^{(i)}:i\in
                         \{1,\dots,n\}\bigr\},&(\ref{eq.A1}.a)\\
        (z_{1}^{(i)},1)&\mbox{ if }\delta(s,x)=z_{1}^{(i)}, i\in
                         \{1,\dots,n\},&(\ref{eq.A1}.b)
      \end{array}\right.
  \end{equation}
  and for $z\in Z_{i}, x\in X$ and $j\ge 1$\
  \begin{equation}\label{eq.A2}
    \hat\delta((z,j),x):= \left\{
      \begin{array}{l@{,}lr}(\delta(z,x),0)
        &\mbox{ if } \delta(z,x)\notin Z_{i}\vee j=\kappa_{i},&(\ref{eq.A2}.a)\\
        (z_{j+1}^{(i)},j+1)&\mbox{ if }j<\kappa_{i}\wedge\delta(z,x)=z_{j+1}^{(i)},
                             \mbox{ and},&(\ref{eq.A2}.b)\\
        (\delta(z,x),j)&\mbox{ otherwise.} ,&(\ref{eq.A2}.c)
      \end{array}\right.
  \end{equation}
  
  Then $\hat\delta((z,j),x)=(\delta(z,x),j')$ with $j'\in \{0,j,j+1\}$, and,
  consequently, $\hat\delta((z,j),w)=(\delta(z,w),j'')$ for $w\in X^{*}$ and
  some $j''$.

  Moreover the following items hold.
  \begin{enumerate}
  \item if $z,\delta(z,x)\in Z_{i}$ and $j< \kappa_{i}$ then
    $\hat\delta((z,j),x)=(\delta(z,x),j')$ where $j'\in \{j,j+1\}$,
    and\label{i.1}
  \item if $\hat\delta((s,j),x)=(s',j+1)$ then
    $s'= \delta(s,x)= z^{(i)}_{j+1}$ for some $i\in\{1,\dots,n\}$.\label{i.2}
  \item[]This implies
  \item If $j\ge1$, $\hat\delta((s,0),w)=(\delta(s,w),j)$ and
    $\delta(s,w)\in Z_{i}$ then
    $\{\delta(s,w'):w'\sqsubseteq w\}\supseteq
    \{z_{1}^{(i)},\dots,z_{j}^{(i)}\}$.\label{i.3}
  \end{enumerate}
  If $\mathrm{Inf}(\hat\mathcal{A}; \xi)\cap T\ne\emptyset$ then there are
  infinitely many prefixes $u\sqsubset\xi$ such that
  $\hat\delta((s_{0},0),u)= (z_{\kappa_{i}}^{(i)},\kappa_{i})$ for some
  $i$. From Item~\ref{i.3} above and
  $\hat\delta((z_{\kappa_{i}}^{(i)},\kappa_{i}),x)=(\delta(z_{\kappa_{i}},x),0)$
  we obtain infinitely many $u_{\ell},w_{\ell}\in X^{*}$ and $x_{\ell}\in X$
  such that
  $u_{\ell}\cdot x_{\ell}\cdot w_{\ell}\sqsubset u_{\ell+1}\sqsubset \xi$
  where $\hat\delta((s_{0},0),u_{\ell})=(z_{\kappa_{i}}^{(i)},\kappa_{i})$,
  $\hat\delta((s_{0},0),u_{\ell}\cdot
  x_{\ell})=(\delta(z_{\kappa_{i}}^{(i)},x_{\ell}),0)$ and
  $\{\delta(z_{\kappa_{i}}^{(i)},x_{\ell}\cdot w): w\sqsubseteq
  w_{\ell}\}\supseteq Z_{i}$.

  Since $Z_{i}$ is a strongly connected component,
  $\mathrm{Inf}(\mathcal{A}; \xi)=Z_{i}$ and
  $\xi\in L(\mathcal{A},\mathcal{T})$ follow.

  Let now $\mathrm{Inf}(\hat\mathcal{A}; \xi)\cap T=\emptyset$, and assume
  $\mathrm{Inf}(\mathcal{A}; \xi)=Z_{i}$ for some $i$. Since
  $\mathrm{Inf}(\hat\mathcal{A}; \xi)\cap T=\emptyset$, there is a
  $u\sqsubset \xi$ such that
  $\{\delta(s_{0},w):u\sqsubseteq w\sqsubset \xi\}\subseteq Z_{i}$ and
  $\{\hat\delta((s_{0},0),w):u\sqsubseteq w\sqsubset \xi\}\cap
  T=\emptyset$. Then for all
  $\hat\delta((s_{0},0),w)=(\delta(s_{0},w),j_{w}),u\sqsubseteq w\sqsubset
  \xi,$ we have $j_{w}<\kappa_{i}$. Let
  $\hat j:= \max\{j_{w}: u\sqsubseteq w\sqsubset \xi\}$. Fix
  $\hat w, u\sqsubseteq \hat w\sqsubset\xi$ such that
  $\hat \delta((s_{0},0),\hat w)=(\delta(s_{0},\hat w),\hat j)$. Then
  Item~\ref{i.1} implies $\hat \delta((s_{0},0),w)=(\delta(s_{0},w),\hat j)$
  for all $w, \hat w\sqsubseteq w\sqsubset\xi$. Thus, in view of
  Item~\ref{i.2},
  $z_{\hat j+1}^{(i)}\notin \{\delta(s_{0},w):\hat w\sqsubseteq
  w\sqsubset \xi\}$ and, consequently, $z_{\hat j+1}^{(i)}\notin Z_{i}$ which
  contradicts $\mathrm{Inf}(\mathcal{A}; \xi)=Z_{i}$.
\end{proof}

\section{Algorithms}
\label{sec.implement}

After the prerequisites derived in the preceding sections we present the
polynomial constructions. To this end we refer to classical graph algorithms
(cf. \cite{Cormanetal,BaaseVanGelder}) and to the results of
\cite{SelWa08}. It is well known that the strongly connected components of a
graph can be estimated in polynomial time of the size of the graph. Similarly,
its condensation graph, that is, the graph having as vertices the strongly
connected components and as edges the connections induced by the underlying
graph, can be also constructed in polynomial time. This allows also to find
the terminal strongly connected components $\mathrm{SCC}_{\mathcal{A}}^{t}$ in
polynomial time.

Moreover we have the following.
\begin{lemma}\label{l.SW}\mbox{\upshape{}\cite{SelWa08}} Let
  $\mathcal{A}=(X,S,s_0,\delta)$ be deterministic automaton. Then, for $Z,Z'\subseteq S$,  the
  predicates $Z\subseteq Z'$ and $Z\in \mathrm{LOOP}_{\mathcal{A}}$ 
  are decidable in non-deterministic logarithmic space.
\end{lemma}

\subsection{Muller automata for the Automatic Baire property}
\label{sec.Mul}
We show that the equation
\begin{equation}
  \label{eq.L}
L(\mathcal{A},\mathcal{T})\ \Delta\
L(\mathcal{A}_{1},\mathcal{T}_{1})\subseteq X^{\omega}\setminus
L(\mathcal{A}_{2},\mathcal{T}_{2})
\end{equation} is satisfied for automata
$\mathcal{A}_{1},\mathcal{A}_{2}$ and tables $\mathcal{T}_{1},\mathcal{T}_{2}$
which can be constructed in polynomial time from $\mathcal{A}$ and
$\mathcal{T}$. We have here to construct the complement of
the first Baire category set
$L\bigl(\mathcal{A},2^{S}\setminus\mathrm{SCC}_{\mathcal{A}}^{t}\bigr)$ in
order to avoid a possible exponential blow-up (cf. \cite{dlt19/boker}) in the
size of the table
$(2^{S}\setminus\mathrm{SCC}_{\mathcal{A}}^{t})\cap
\mathrm{LOOP}_{\mathcal{A}}$.

According to Eq.~(\ref{eq.fin}) and Lemma~\ref{l.boolop} we define
$\mathcal{A}_{2}:=\mathcal{A}$ and
$\mathcal{T}_{2}:=\mathrm{SCC}_{\mathcal{A}}^{t}$.

The definition of $\mathcal{A}_{1}:=\mathcal{A'}$ and $\mathcal{T}_{1}:=
\{s_Z:Z\in \mathcal{T}\cap\mathrm{SCC}_{\mathcal{A}}^{t}\}$ is given 
in Corollary~\ref{c.open}.

Then an algorithm constructing $\mathcal{A}_{1},\mathcal{A}_{2}$ and
$\mathcal{T}_{1}, \mathcal{T}_{2}$ from $\mathcal{A}=(X,S,s_0,\delta)$ and
$\mathcal{T}\subseteq 2^{S}$ proceeds in the following way.
\begin{enumerate}
  \item Detect the strongly connected components $\mathrm{SCC}_{\mathcal{A}}$
    of the automaton $\mathcal{A}$.\label{alg1}
  \item Construct the condensation graph
    $(\mathrm{SCC}_{\mathcal{A}},\mapsto)$.\label{alg2}
  \item Identify, in $(\mathrm{SCC}_{\mathcal{A}},\mapsto)$, the terminal
    strongly connected components
    $\mathrm{SCC}_{\mathcal{A}}^{t}$.\label{alg3}
  \item Compare $\mathrm{SCC}_{\mathcal{A}}^{t}$ with $\mathcal{T}$ to obtain
    $\mathrm{SCC}_{\mathcal{A}}^{t} \cap\mathcal{T}$.
  \end{enumerate}
  Items \ref{alg1},\ref{alg2} and \ref{alg3} are well-known graph algorithms
  polynomial in $\mathrm{size}(\mathcal{A})$. For the comparison of
  $\mathrm{SCC}_{\mathcal{A}}^{t}$ with $\mathcal{T}$ we use Lemma~\ref{l.SW}
  to check whether $Z\subseteq Z'$ and $|Z|=|Z'|$ for
  $Z\in \mathrm{SCC}_{\mathcal{A}}^{t}$ and $Z'\in \mathcal{T}$. This can be
  done in polynomial time in
  $\mathrm{size}(\mathcal{A})+\mathrm{size}(\mathcal{T})$.
  
\subsection{B\"uchi automata for the automatic Baire property}
\label{sec.Bue}

In order to obtain the B\"uchi automata $(\mathcal{B}_{1},T_{1})$ and
$(\mathcal{B}_{2},T_{2})$ we use the Muller automata
$\mathcal{A}_{1}, \mathcal{A}_{2}$ and tables
$\mathcal{T}_{1}, \mathcal{T}_{2}$ satisfying Eq.~(\ref{eq.L}). For
$(\mathcal{A}_{1},\mathcal{T}_{1})$ the transformation to
$(\mathcal{B}_{1},T_{1})$ is straightforward. Set
$\mathcal{B}_{1}:= \mathcal{A}_{1}$ and
$T_{1}:= \mathcal{T}\cap\mathrm{SCC}_{\mathcal{A}_1}^{t}$. This results in a
so-called weak B\"uchi automaton $(\mathcal{B},T)$ where either $R\subseteq T$
or $R\cap T=\emptyset$ for $R\in \mathrm{LOOP}_{\mathcal{B}_1}$.

The table $\mathcal{T}_{2}=\mathrm{SCC}_{\mathcal{A}_2}^{t}$ of the Muller
automaton $(\mathcal{A}_{2},\mathcal{T}_{2})$ satisfies the hypothesis of
Theorem~\ref{th.maxloop}. Thus we can construct the B\"uchi automaton
$(\mathcal{B}_{2},T_{2})$ according to the construction in the proof. Since
the size of $(\mathcal{B}_{2},T_{2})$ is only quadratic in the number of
states of $\mathcal{A}_{2}$, this can be done in polynomial time.

% \bibliography{cie_staiger}%
% \bibliographystyle{alpha}

\end{document}